\documentstyle[11pt,paspconf]{article}
\begin{document}
\title{Observations and Theory of Dynamical Triggers for Star Formation}
\author{B. G. Elmegreen}
\affil{IBM Research Division, T.J. Watson Research Center, Yorktown
Heights, NY 10598}

\begin{abstract}
Star formation triggering mechanisms are reviewed, including
the direct compression of clouds and globules, the compression
and collapse of molecular clouds at the edges of HII regions and
supernovae, the expansion and collapse of giant rings and shells in 
galaxy disks, and the collision and collapse between two clouds. 
Collapse criteria are given. A comprehensive tabulation
of regions where these four types of triggering have been found
suggests that dynamical processes sustain and amplify a 
high fraction of all star formation that begins spontaneously in 
normal galaxy disks. 
\end{abstract}


To be published in the proceedings of the NASA Conference, 
"Origins of Galaxies, Stars, Planets and Life,"
ed. C.E. Woodward, H.A. Thronson, \& M. Shull, 
Astronomical Society of the Pacific Conference Series, 1998.

\section{Introduction}

Star formation requires dense, self-gravitating
gas, so 
compression is often considered to be a 
requisite, or helpful, precursor.
The earliest models of triggered star formation
were not concerned with gas density 
however, but with gas and star
motions: Opik (1953) and Oort (1954)
proposed that the expansion of OB associations observed by Blaauw 
\& Morgan (1953)
was the result of star formation in moving gas, expelled from a 
supernova or stellar cluster by high pressures. 
Today, we think
that the expansion of OB associations, aside from runaway O stars, 
is the result of gravitational
unbinding of the young embedded stellar cluster after the gas leaves
(Zwicky 1953; see review in Lada 1991), so the original 
motivation for triggering has changed. 

The idea that high interstellar pressures could directly 
squeeze a pre-existing cloud and cause it to collapse to a star
followed from the theory of this process by Ebert (1955) and Bonner (1956),
and was part of the triggering scenario applied to dark globules 
by Dibai (1958).  Star formation in such globules had previously been
considered by Bok \& Reilly (1947) and others.
This process is now thought to have widespread application to bright rims, 
cometary globules, globules in HII regions, and isolated
clouds subject to high pressures from supernovae and other 
disturbances (see review in Klein, Whitaker \& Sandford 1985).

Two other observations that led to early suggestions of triggering 
concerned spatial progressions of star formation. 
For OB associations, 
Blaauw (1964) observed a spatial separation of subgroups with different ages,
and this 
led to the idea that star formation is triggered sequentially 
in a molecular cloud
by pressures from HII regions (Elmegreen \& Lada
1977) and supernovae
(van Till, Loren, \& Davis, 1975;
\"Ogelman \& Maran 1976; Herbst \& Assousa 1977). 
Larger scale models of triggering 
(Dopita, Mathewson \& Ford 1985) similarly followed from 
an age progression in the Constellation III region of the
Large Magellanic Clouds 
(Westerlund \& Mathewson 1966).
(This particular age progression is now in doubt; see
Braun et al. 1997).
Observations of age progressions are the basis 
for a second triggering scenario in which relatively low density gas 
is 
moved and compressed by a shock front until it
collapses gravitationally.
The collapsed gas forms dense molecular cores in which star
formation occurs at a high rate, possibly forming a dense cluster.

A third suggestion for triggering came 
from models of gravitational instabilities during
shock compression (Woodward 1976) and from the observation of star
formation at the interface between an expanding
HI shell and a dense cloud (Loren 1976). 
This led Loren to the cloud-collision model, even though 
he recognized that 
cloud collisions could be destructive (Stone 1970).
A second example, LkH$\alpha$, was soon found (Loren 1977), and then 
Dickel et al. (1978) applied the model to the W75/DR 21 region.
Since then, numerous models (e.g., Kimura \& Tosa 1996)
have shown how collisions can lead to gravitational
instabilities in the shocked region, and
recent observations of this affect have been reported by 
Hasegawa et al. (1994) and others. 

Evidently, there are three distinct 
triggering mechanisms that are commonly discussed: 
(1) direct compression of pre-existing globules or density enhancements in 
a cloud ("globule squeezing"), (2) accumulation of 
gas into a dense ridge or shell that collapses gravitationally
into dense cores ("collect and collapse"), and
(3) cloud collisions. 

Another early line of evidence for triggered star formation
was the observation of abundance anomalies 
of short-lived radio isotopes in meteorites (Lee, Papanastassiou, \&
Wasserburg 1977).
Cameron \& Truran (1977)
suggested this implied the solar system was triggered by a supernova.
Recent theoretical studies of such triggering 
in the globule-squeezing scenario were made by 
Ramadurai (1986), Boss (1995), Cameron, et al. (1995), and
Valhala, Cameron, \& Hoflich, (1996).
Boss considered slow stellar winds for the triggering, and the
others considered supernovae. 

These triggering mechanisms may also have applications to 
star formation in galactic spiral arms. The need for such triggering was  
first discussed in the context of the density wave theory by
Roberts (1969) when it was thought that underlying spiral waves
were weak and most of the appearance of spiral structure was from 
triggered star formation. Today, infrared observations 
(Elmegreen \& Elmegreen 1984; Rix \& Rieke 1993; Block et al. 1994)
reveal strong arms in even the old stellar disk, 
and H$\alpha$/CO arm-interarm contrasts 
suggest only moderate amounts of triggering, or, perhaps
no triggering in some cases
(Garcia-Burillo, Guelin, \& Cernicharo 1993; Sempere \& Garcia-Burillo 1997).
Nevertheless, the globule-squeezing mechanism
has been applied to spiral arms by Woodward (1976), the 
collect-and-collapse mechanism by Elmegreen (1979a, 1994a), 
Balbus \& Cowie (1985), Tomisaka (1987), Balbus (1988), and others, 
and the cloud-collision mechanism by 
Kwan \& Valdes (1983), Scoville, Sanders, \& Clemens (1986) 
and others. 

At the present time, 
there is no evidence in high resolution dust maps of 
in M51 (Block et al. 1997) for the comet-shaped, shocked  
clouds
that were predicted by Woodward (1976) to trigger star formation 
in spiral arms. Indeed,  
the large interarm clouds tend to be
spiral-shaped rather than spherical (Block et al. 1997). 
There is also no direct evidence for cloud collision-induced star
formation in spiral arms, although this would be very difficult to
obtain with present-day resolution. 
Instead, 
extensive observations of regularly-spaced, giant HI (Elmegreen \& Elmegreen 
1983, 1987; Boulanger \& Viallefond 1992) 
and CO (Rand \& Kulkarni 1990; Rand 1993) 
clouds in spiral arms, combined with observations of a dominant scale
for star formation that has the same size, i.e., that of Gould's Belt 
(Efremov 1995), plus 
observations of a gravity-sensitive criterion for the onset of galactic star
formation (Kennicutt 1989), all
suggest that the dominant star formation process 
in spiral galaxies is a spontaneous gravitational  
instability in the arms and disks.
The same appears to be true in some starburst rings,
because most star formation takes place in giant hot-spots
having a size or separation equal to the local Jeans length 
(Elmegreen 1994b).  Thus {\it spontaneous processes probably dominate the 
onset of star formation on a galactic scale, but
triggered star formation
sustains, amplifies, and disperses what large-scale
instabilities begin.}  

Spiral wave and other galactic-scale 
triggering will not be discussed further here; a comprehensive review
is in Elmegreen (1995).  
A review of all aspects of triggered star formation is in 
Elmegreen (1992). 
Here we cover recent observations and some of the most basic aspects of
the theory. 

\section{Observations of triggered star formation}

There are numerous observations of star formation in 
OB associations and other high pressure regions 
that are thought to be the result of triggering.
Tables 1-4 in the next sections
list some these observations by region, going back about 10 years.
This list was generated by a survey of the literature, with help
from the ADS abstract service, so it may miss some references
that are not in this service or
that do not specifically discuss this topic in the 
abstract.

There is some attempt to organize the list into the proposed mechanisms
of triggering, with a distinction made between small, intermediate, 
and large scales in the following sense: {\it Small scale triggering} 
(Table 1):
direct squeezing of pre-existing clouds or globules by high pressure 
that nearly surrounds the whole cloud. This includes triggering in bright
rims, proplyds, and small cometary globules. {\it Intermediate
scale triggering} (Table 2): compression of a nearby cloud from one side, 
leading to  
a dense ridge of moving gas that 
presumably collapses or recollects into denser
cores in which star clusters eventually form. {\it Large scale triggering}
(Table 3):
accumulation of gas into an expanding shell or ring partially surrounding the
pressure source, with star formation 
in the shell or ring presumably triggered by gravitational 
collapse of swept-up gas. These three scales follow from a comparison
between the size of the trigger{\it ed} 
region (small, medium and large) and the size
of the trigger{\it ing} region. 

The observations suggest that 
triggering occurs in most of the large star-forming 
regions near the Sun. Indeed, triggering of one type or another
may occur inside and around {\it most}
of the clouds in which OB stars form.
For triggering on small and intermediate scales, the extent of triggering
is probably 
limited to the pre-existing dense gas, i.e., to the molecular
clouds in which the first generation
of stars formed.
For large scales, triggering apparently occurs in ambient gas that
was not previously part of the star-forming cloud. Whether all
star formation, including the first generation inside molecular
clouds (e.g., by stray supernovae)
and the formation of the molecular cloud themselves, has been 
triggered by a previous generation of stars is unknown. 
The evidence for galactic-scale gravitational instabilities
discussed in the Introduction suggests that only a small fraction
of star formation is triggered by a previous generation of stars, 
perhaps less than 50\%. Nevertheless, 
if a high fraction, e.g., 99\%, of all cloud and 
star formation is triggered, then 
there will be important implications for spiral structure in 
galaxies (Seiden \& Schulman 1990; Jungwiert \& Palous 1994).

\section{Theory of triggering}

\subsection{Small Scale Triggering: Globule-Squeezing}

\subsubsection{Overview:}

\begin{figure}
\vspace{2.2in}
\includegraphics{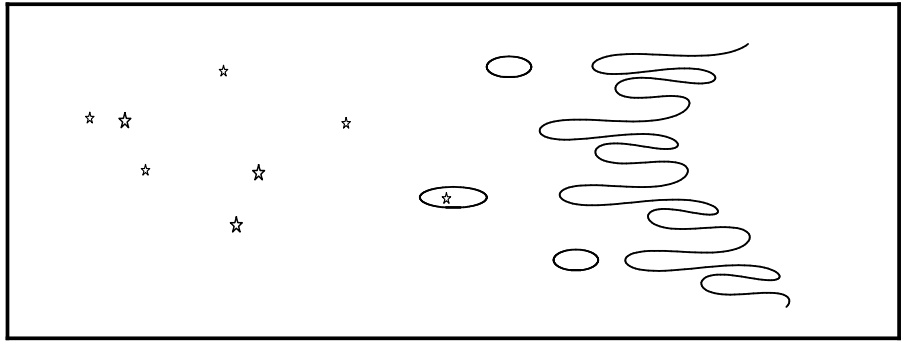}
\caption{Schematic diagram of a young cluster (left) interacting with 
the clumpy structure at the edge of a molecular cloud (right). Some
of the clumps are squeezed into gravitational collapse, forming stars.}
\label{fig:tenfrac}
\end{figure}

Numerous observations of clumpy structure inside and on the periphery
of molecular clouds (Stutzki et al. 1988; 
Falgarone, Phillips \& Walker 1991) suggest that 
young HII regions begin their expansion 
by interacting with cloud clumps.
The high pressures of the HII regions
can squeeze these clumps and trigger star formation. A schematic diagram
is shown in Figure \ref{fig:tenfrac} (Figs. 1-4 are from Elmegreen 1992).

Dense neutral clouds at the edges of HII regions may also be {\it formed}
there, independent of pre-existing molecular cloud clumps.  
Dynamical instabilities in the swept-up gas between
the ionization front 
and the molecular cloud can fragment the material into small 
pieces even without self-gravity. One instability occurs if
the HII region becomes brighter with time (Elmegreen \& Elmegreen 
1978a). Another is the Rayleigh-Taylor instability 
(Schneps, Ho \& Barrett 1980, Schwartz 1985; 
Capriotti 1996), and still others 
result from transverse flows in the thin swept-up
layer 
(Giuliani 1979; Vishniac 1983). Recent applications of the
latter 
instability to expanding HII regions were made by Garcia-Segura
\& Franco (1996).

Numerical simulations of the formation of bright rims and other peripheral
structures are in Bedijn \& Tenorio-Tagle (1984),
Sandford, Whitaker, \& Klein (1982, 1984), 
Klein, Sandford, \& Whitaker (1983), 
Lefloch \& Lazareff (1994), and Elmegreen, Kimura, \& Tosa (1995).
Analytical work on the structure of embedded globules and
cometary clouds is in many references, including
Oort \& Spitzer (1955), Kahn (1969), Dyson (1973), 
Brand (1981),
Bertoldi (1989), and
Bertoldi \& McKee (1990).

The direct observation of neutral globules in HII regions had a difficult
start. 
Dyson (1968) proposed they existed in order to 
power the observed HII turbulence, but it was not until 10 years later that 	
Laques \& Vidal (1978) found them in Orion.
Shaver et al. (1983) then 
observed clumpy ionized structure in a large sample of 
HII regions,
and
Felli et al. (1984) found bright HII emission peaks inside the M17 nebula.
Further studies of Orion by 
Garay, Moran, \& Reid (1987) 
and Churchwell et al. (1987) showed convincingly that some of
the embedded bright rims
are from neutral globules.  

\begin{table}
\caption{Star Formation in Cometary Globules and
Bright Rims} 
\begin{center}\scriptsize
\begin{tabular}{ll}
Region&Reference\\
\tableline

Gum nebula&Brand et al. 1983; Reipurth 1983; Sahu et al. 1988;\\
&Harju et al. 1990; Bhatt 1993; Sridharan 1992;\\ 
&Reipurth \& Pettersson 1993; Vilas-Boas, Myers;\\
& \& Fuller 1994; Gonz\'alez-Alfonso, Cernicharo,\\
& \& Radford 1995; Bourke et al. 1995; Sridharan, Bhatt,\\
& \& Rajagopal 1996; Schoeller et al. 1996\\

Orion &Sugitani et al. 1989; Ramesh 1995; O'Dell et al. 1993,\\
& 1994; McCullough et al. 1995; Cernicharo et al. 1992\\

IC 1396 &Kun, Balazs, \& Toth 1987; Balazs, \& Kun 1989;\\
&Nakano et al 1989; Sugitani et al. 1989; Duvert et al. \\
&1990; Serabyn, G\"usten, \& Mundy 1993; Patel et al.\\
& 1995; Weikard et al. 1996; Moriarty-Schieven, Xie,\\
& \& Patel 1996; Saraceno et al. 1996\\

Rosette&Block 1990;
Gonz\'alez-Alfonso, \& Cernicharo 1994;\\
&Patel, Xie, \& Goldsmith 1993;
Indrani \& Sridharan 1994;\\

Ophiuchus&De Geus 1992\\
L810&Neckel \& Staude 1990\\
L1206&Sugitani et al. 1989; Ressler, \& Shure 1991\\
L1780&Toth et al. 1995\\
L1582&Zhou, Butner, \& Evans 1988\\

IC 1805&Heyer et al. 1996\\
IC 1848&Loren \& Wootten 1978; Lefloch \& Lazareff 1995\\
IC4628&King 1987\\

NGC 281&Megeath \& Wilson 1997\\
NGC2264&Tauber, Lis, \& Goldsmith 1993\\
NGC5367&White 1993\\

M8&Caulet 1997\\
M16&Hester et al. 1996\\ 

G10.6-0.4, W33&Ho, Klein, \& Haschick 1986\\
G110-13&Odenwald et al. 1992\\

Thumbprint glob.&Lehtinen et al. 1995\\

Carina&Ogura \& Walsh 1992; Megeath et al. 1996\\
BD +40$^\circ$4124&Hillenbrand et al. 1995\\
\end{tabular}
\end{center}
\end{table}

High resolution observations
with {\it Hubble Space Telescope} now find solar-system size globules in
Orion (Hester et al. 1991; O'Dell, et al. 1993, 1994),
M16 (Hester et al. 1996), and M8 (Caulet 1997). 
Other similar regions have been found in the Sco-Cen association 
(Bertoldi \& Jenkins 1992).
Some of these neutral regions may be protostellar disks, 
as argued in the papers on Orion, but Hester et al. (1996)
claims that many are evaporating globules
rather than disks. 
This follows from the short exposure times of the neutral regions, 
as determined from their proximity to the rest of the molecular cloud
in M16.

Recent evidence for star formation in embedded globules is
summarized in Table 1.
Young embedded or adjacent stars are connected with these globules, but 
it is uncertain whether this star formation was triggered by compression
or was there before. For example, 
star formation could have occurred in the dense gas
independently of the ionization and simply been exposed when the ionization 
cleared away the peripheral gas. This situation was discussed
in some detail by Elmegreen (1992), and a schematic diagram of it is in
Hester et al. (1996).

Dense neutral clouds can also become engulfed by high pressure from
more distant sources, such as supernovae or stellar winds, or perhaps
from older HII regions that have already expanded. 
These make comet shapes when the pressure is one-sided.
Catalogs of such cometary globules are in 
Hawarden \& Brand (1976), Sandqvist (1976), Reipurth (1983), and
Zealey et al. (1983).
Observations of star formation in cometary globules are also included in 
Table 1. 

The observations suggest there are three distinct morphologies for
dense neutral structures exposed to high pressures:
(1) they may be isolated neutral globules that are either protostellar
disks or pressurized cloud pieces, (2) they may be cometary or elephant
trunk globules, which contain a dense cloud at the brightest end
of an elongated structure
and a connected neutral tail at the other end,
and (3) and they may be bright rims, which are like 
cometary globules but with very short tails, 
or no tails, connecting the dense head
to the rest of the cloud. 

Catalogs of bright-rimmed clouds with embedded star
formation are in Sugitani, Fukui, \& Ogura (1991), 
Sugitani \& Ogura (1994), and
Indrani \& Sridharan (1995).
Other recent observations are in Table 1. 
Sugitani, Tamura \& Ogura (1995) found small-scale age
sequences inside bright rimmed clouds. Similar small scale
sequences were found in the Orion
ridge by Chandler \& Carlstrom (1996).

The interaction between a supernova
or other shock and a non-magnetic dense cloud was studied numerically by
many authors, including
Woodward (1976), 
Nittmann et al. (1982),
Heathcote \& Brand (1983), 
Tenorio-Tagle \& Rozyczka (1986),
Rozyczka \& Tenorio-Tagle (1987), 
Falle \& Giddings (1989),
Bedogni \& Woodward (1990),
Stone \& Norman (1992),
Klein, McKee, \& Colella (1994), and
Xu \& Stone (1995).
Studies with a magnetic field were made by
Nittmann (1981) and  
Mac Low et al. (1994).
These tend to show compression of the cloud with the simultaneous
formation of a tail of material shred off from
the cloud's surface. 
If the cloud is initially dense, the compression is strong, 
and the compressed gas cools so it
becomes very dense after the compression, then the interior 
can collapse to 
make a star before the surface gets shred away 
by Kelvin-Helmholtz and other
instabilities. 

Observations of the interaction between the Cygnus 
supernova remnant and an ambient cloud are in 
Fesen et al. (1982, 1992), 
Graham et al. (1995), and Levinson et al. (1996).
Observations of the interaction between the remnant W44 and
small embedded clouds 
are in Rho et al. (1994). These interactions seem much too young for
triggering to have happened yet. 
It is not even clear whether a single supernova exploding in the
ambient medium or in the progenitor stars' wind debris can
trigger star formation by itself. Usually supernova 
triggering takes place in an environment that has already
been highly perturbed by a long history of stellar winds and
HII regions, such as the Gum nebula or an OB association. 

\subsubsection{Theory of Triggering in Globules:}

The basic process of triggering is the increased pressure
at the edge of a clump.  The maximum pressure for a stable,
self-gravitating isothermal sphere was given by 
Ebert (1955) and Bonner (1956): 
\begin{equation} P_{max}={{1.4c^8}\over{G^3M^2}}.\end{equation}
Chi\`eze (1987) writes this as
\begin{equation}P_{max}={{xGM^2}\over{R^4}} \;\;{\rm where}\;\;
x={{4-3\gamma}\over{8\pi\gamma}}\end{equation}
for equation of state $P=\rho^\gamma$. 
When the boundary pressure of the sphere, $P$, exceeds $P_{max}$, the sphere
collapses. In the Chi\`eze condition, the sphere is always stable
regardless of $P$ if 
$\gamma>4/3$, because then the pressure force that resists the collapse
increases faster during compression than the gravitational force that
drives it.

This stability at high $\gamma$ also applies to magnetic clouds, 
for which there is an effective $\gamma=2$ for compression perpendicular
to the field lines without diffusion. The reason for this is that
magnetic pressure equals $B^2/(8\pi)$ and $B\propto$ density with
perpendicular compression and flux-freezing. Compression 
perpendicular to $B$ cannot trigger collapse because the internal magnetic
pressure goes up faster than the internal self-gravitational energy density. 
Thus we expect two conditions
for the collapse of both polytropic clouds and isothermal magnetic clouds:
one to limit how fast pressure increases with density, and the other
to specify how much boundary pressure is needed for self-gravity
to overcome internal pressure. 

Mouschovias \& Spitzer (1976) wrote the collapse 
condition for an isothermal sphere with
an initially uniform magnetic field:
\begin{equation} P>P_{max}={{1.89c^8}\over{G^3M^2}}
\left(1-\left[{{M_{mag}}\over{M}}\right]^{2/3}\right)^{-3},
\end{equation}
for $M>M_{mag}$, where
\begin{equation} M_{mag}={{B^3}\over{280G^{3/2}\rho^2}}.
\end{equation}
This second condition is related to the relative contraction
parallel and perpendicular to the field. If $M<<M_{mag}$, then 
a significant fraction of 
the compression-induced motion is perpendicular to the field 
lines and the field pressure increases 
as fast as, or faster than, self-gravity; this
gives absolute stability, as discussed above.
If
$M>M_{mag}$, then there is significant motion parallel
to the field lines during the contraction, driven by self-gravity.
In the latter case, the central density goes up faster than in the
$M<M_{mag}$ case, and gravity can 
overcome internal pressure support when $P>P_{max}$. 

We can invert the $P>P_{max}$ condition 
to write 
\begin{equation}
M>M_{P}={{1.37c^4}\over{G^{3/2}P^{1/2}
\left(1-\left[M_{mag}/M\right]^{2/3}\right)^{3/2}}}.
\end{equation}
Then the Tomisaka et al. (1989) condition with rotation and\
magnetic fields is 
\begin{equation}
M>M_{crit}=\left(M_P^2+\left[{{4.8cj}\over{G}}\right]^2\right)^{1/2}
\end{equation}
for specific angular momentum $j$ parallel to the field.

Torsional magnetic fields decrease the critical mass because
they pinch together the poloidal field lines, like an extra
gravitational force; this implies that a torsional magnetic wave 
may trigger collapse in a marginally stable cloud (Tomisaka 1991).
Tomisaka et al. (1989) point 
out that there are two stable solutions for given
magnetic flux, mass and angular momentum, corresponding to low-density,
low-spin, and 
pressure-bound, or high-density, high-spin, and gravity-bound. 
Tohline \& Christodoulou (1988) discussed two similar solutions
for the non-magnetic case. 

Cooling of gas at higher density also decreases stability, in 
the sense that much milder pressure fluctuations are required to
trigger collapse when the adiabatic index $\gamma$ is small
(Tohline, Bodenheimer, \& Christodoulou 1987).

The parameters occurring in these equations are 
boundary pressure $P$, magnetic field strength, $B$, 
density, $\rho$, 
mass, $M$, velocity dispersion, $c$, and specific angular momentum 
$j$.  They should all be evaluated before 
before any self-gravitational contraction
occurs, i.e., they are the initial conditions for the globule.
In the case
of triggering, they are the conditions after the globule
is squeezed, but before it collapses. The $P$, for example, 
is the pressure in the HII region surrounding the neutral cloud, and
$B$ is the magnetic field strength in the compressed globule, which can 
be larger than in the pre-compressed molecular cloud. 

These equations imply 
that if a spheroid is stable before compression because
\begin{equation} M<M_{mag}={{B^3}\over{280G^{3/2}\rho^2}},
\end{equation}
then it will be even more stable after compression
{\it perpendicular} to the field lines because 
$B\propto\rho$ in this case. 
Conversely, 
if a stable region of mass $M<M_{mag}$ is compressed
{\it parallel} to the field, where $B\sim$ constant and $\rho$
increases, then $M_{mag}$ decreases giving $M>M_{mag}$, and the
condition for instability is essentially the field-free one: $P>P_{max}$.  
Thus direct compression triggers instability only
if a substantial part of it is parallel to the magnetic field
or the field is initially weak. 

When there is magnetic diffusion, the situation changes. 	
The diffusion rate is
\begin{equation}\tau_{diff}={{\nu_{in}}\over{\omega_A^2}}\propto
{{n_{i}\sigma_{in}c}\over{k^2v_A^2}}\propto{{n_i\rho R^2}\over
{B^2}}\propto{{\rho^{3/2}R^2}\over{B^2}}
\end{equation}
for ion-neutral collision rate $\nu_{in}$, Alfven wave frequency
$\omega_A$, ion density $n_i$, ion-neutral collision cross section 
$\sigma_{in}$, Alfven wavenumber $k$, Alfven speed $v_A$, and
size of region $R$; 
$n_i$ is assumed to be proportional to 
$\rho^{1/2}$ (Elmegreen 1979b).
Compared to the free-fall rate, the diffusion rate is
\begin{equation} 
{{\tau_{diff}}\over{\tau_{ff}}}\propto
\left({{\rho R}\over{B}}\right)^2 .
\end{equation}
For compression of a fixed mass {\it parallel} to ${\bf B}$, 
both $\rho R$ and $B$ are constant, so $\tau_{diff}/\tau_{ff}$
is constant and the compression does not speed up the diffusion 
relative to the collapse.
For compression {\it perpendicular} to ${\bf B}$,
$\rho/B$ is constant and $R$ decreases, leading to more rapid
diffusion. 

\begin{figure}
\vspace{3.0in}
\includegraphics{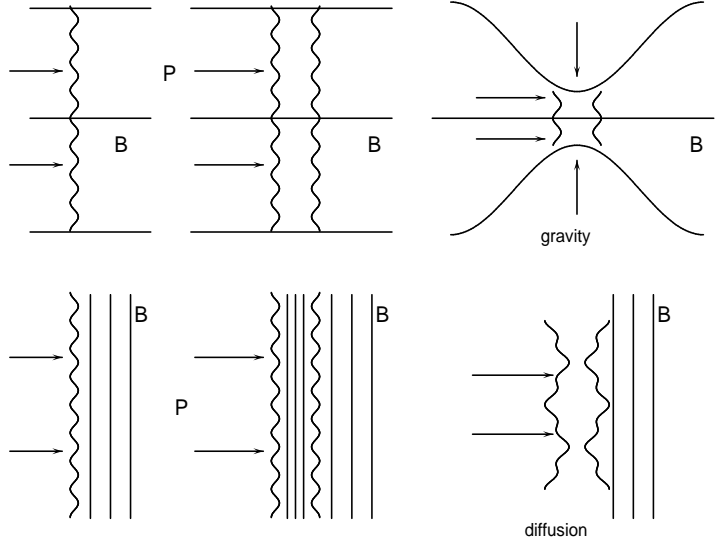}
\caption{Schematic diagram of compressions parallel (top) and
perpendicular to the magnetic field with the collapse to 
star formation resulting from gravity overcoming internal magnetic 
pressure and from magnetic diffusion, respectively.} 
\label{fig:tenpar}
\end{figure}

Thus we have two cases:\\
(1) compression of a stable region 
parallel to ${\bf B}$ may trigger collapse by decreasing $M_{mag}$,
but it can make magnetic diffusion relatively slow.\\
(2) compression of a stable region 
{\it perpendicular} to ${\bf B}$ makes the
region dynamically more stable at first, 
but collapse can follow quickly after the field diffuses away.

It is illustrative to 
write $\rho$ and $B$ in terms of $P$ using $\alpha=
\rho c^2/P$ and $\beta=B^2/(8\pi P)$.
Then the two conditions for the
collapse are
\begin{equation}
M>M_{\rm mag} = 0.011 {\it
M}_{\odot}{{c_{0.1}^4\beta^{3/2}}\over{P_6^{1/2}\alpha^2}}\end{equation}
and
\begin{equation}
M>M_{\rm P} = 0.034 {\it
M}_{\odot}{{c_{0.1}^4}\over{P_6^{1/2}}}\left(1-(M_{\rm mag}/M)^{2/3}
\right)^{-3/2}.\end{equation}
Here, $c_{0.1}$ is the velocity dispersion in units of
0.1 km s$^{-1}$, and $P_6$ is the pressure in units of 
$10^6k_B$ for Boltzmann constant $k_B$. 
As before, the second condition on pressure has been written in terms of a
critical mass.   If the magnetic
and thermal pressures in the globule are comparable to the external
pressure, then these two conditions are about the same.
 
Evidently, 
compression with a
magnetic field is very different from compression without a field.
Pressure alone does not trigger dynamical collapse when a magnetic
field is present unless the field is weak enough to make
$M>>M_{mag}$ initially. 
For compression predominantly parallel to the
field, collapse is triggered because $M_{\rm mag}$ decreases during
the compression and gravity overcomes magnetism.
For compression predominantly perpendicular to the field,
dynamical collapse is inhibited because of the increased field
strength, but the field diffuses out of the gas more rapidly.  Then
enhanced diffusion is what triggers collapse, by lowering $M_{\it
mag}$ to a value less than $M$.
A schematic diagram of these two situations is shown in Figure 
\ref{fig:tenpar}.
Of course, compression parallel to the field can be followed by 
collapse perpendicular to the field, and vice versa, and after 
both motions, the cloud can be unstable {\it and} have rapid diffusion.

The 
spontaneous formation and collapse of 
globules inside filaments is another way to form stars. 
Filaments are very common in the interstellar medium. They 
may form by magnetic processes 
(Elmegreen 1994d) or by the gravitational collapse of shocked
layers (Miyama, Narita, \& Hayashi 1987).
A filament whose support in the transverse direction is strongly dominated
by magnetic forces will collapse to stable oblate globules, while one
with less initial magnetic energy will collapse to unstable globules
and eventually stars.  
The threshold between these two cases occurs for an initial
ratio of thermal to magnetic pressure equal to 0.02 (Tomisaka 1995).
The fragmentation of a cylinder is strongly influenced by geometry, 
giving a characteristic separation between globules that scales with 
the filament width, rather than the Jeans length for sufficiently small
Jeans length (Bastien et al. 1991). This is in agreement
with the observed regular spacing between globules in dark 
filaments (Schneider \& Elmegreen 1979) and spiral arms (Elmegreen 
\& Elmegreen 1983).
Numerical simulations of the collapse of a filament 
to the stage
where dense protostellar disks form are in Nakamura et al. (1995)
and Tomisaka (1996). 
The collapse and fragmentation of prolate clouds was considered by 
Nelson \& Papaloizou (1993) and Bonnell \& Bastien (1993).

The stability of interstellar clouds to outside fluctuations in
pressure, radiation, and other variables is
no doubt more complicated than these simple analyses suggest. 
Internal motions, turbulence, magnetic waves, fragmentation, heating
and cooling, ionization, and other 
processes can alter the conditions and outcome of triggered star formation. 
Some of these processes are included in the many published investigations on
cloud collapse. Even without magnetic fields or rotation, there has been a
considerable amount of analysis on instability and the collapse phase, as in 
Larson (1969), Penston (1969),
Shu (1977), Hunter (1977), 
Stahler, Shu, \& Taam (1980),
Whitworth \& Summers (1985), 
Blottiau, Chieze, \& Bouquet (1988),
Ori \& Piran (1988), 
Suto \& Silk (1988),
Foster \& Chevalier (1993),
Pen (1994),
Boily \& Lynden-Bell (1995),
Tsai \& Hsu (1995), and
Whitworth et al. (1996).
The results of the most recent of these 
studies indicate that the initial and collapse
phases of star-forming clouds 
cannot be well represented by solutions for a singular 
isothermal spheres, and that
mild kinematic disturbances outside and inside the initial cloud can have
important consequence for cloud evolution and collapse. This means that 
the trigger for some collapses can be very subtle.  

Collapse studies including rotation usually investigate binary or multiple
star formation; they are rarely concerned with triggering.  
Nevertheless, a few examples are useful for comparison, such as
Boss \& Black (1982),
Terebey, Shu, \& Cassen (1984),
Myhill \& Kaula (1992),
Bonnell \& Bastien (1992),
Burkert \& Bodenheimer (1993),
Boss (1993),
Sigalotti \& Klapp (1994),
Boss, \& Myhill (1995),  and
Boss (1996).
It is not clear whether the multiplicity of stars is affected by
triggering. If triggering by one-sided compression tends to make
flattened clouds, then multiplicity may be increased
compared to the formation of stars in centrally-condensed,
spherical clouds (Chapman et al. 1992). 

The collapse of magnetic clouds, with or without rotation, also has
unknown implications for triggering. Usually the collapse is taken
to begin from 
a near-static initial condition, as in the Mouschovias \& Spitzer problem 
reviewed above. If the collapse is initiated by an implosion,  
or a torsional magnetic wave, or some other fluctuation in the 
environment,  
then the conditions for star formation could change. 
Variations in ionizing radiation have been considered by McKee (1989), 
for example. 
The collapse of magnetic clouds, usually following ambipolar diffusion, 
has been considered recently by  
Galli \& Shu (1993a,b),
Fiedler \& Mouschovias (1993),
Basu \& Mouschovias (1994, 1995a,b),
Ciolek \& Mouschovias (1994, 1995), and
Boss (1997).

\subsection{Intermediate scale triggering: Collect and Collapse}
\label{sec:intermed}

As an HII region ages, its expansion moves more and more of the 
neutral cloud,
accumulating it into a dense neutral ridge at 
the nebula/cloud 
interface. 
This ridge may be unstable 
on a short, internal-crossing timescale because of kinematic, magnetic,
and gravitational 
processes (Elmegreen \& Elmegreen 1978b; 
Giuliani 1979; Doroshkevich 1980; 
Welter \& Schmid-Burgk 1981; Welter 1982; Vishniac 1983, 1994; 
Bertschinger 1986; 
Vishniac \& Ryu 1989; Wardle 1991; 
Nishi 1992; Yoshida \& Habe 1992; Kimura \& Tosa 1991, 1993;
Lubow \& Pringle 1993;
Mac Low \& Norman 1993; 
Strickland \& Blondin 1995; Garcia-Segura \& Franco 1996),
and this fast instability may be oscillatory or monotonic, depending
on the deceleration. 

\begin{figure}
\vspace{2.2in}
\includegraphics{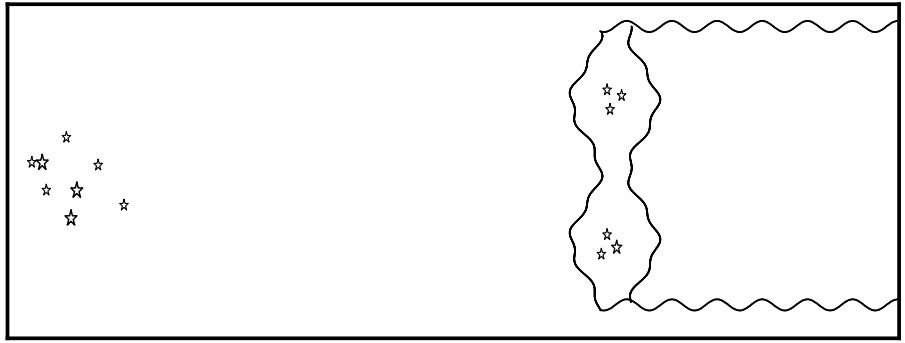}
\caption{Schematic diagram of the motion and compression
of neutral gas at the edge of an expanded HII region, showing a
collapse into two large clumps and young stars inside.} 
\label{fig:teninter}
\end{figure}

However, 
because the layer is thin in one dimension, it is also
unstable on a longer timescale to 
gravitational 
collapse along its length. Gravity always acts to make the
geometry of a cloud more spherical if it is pressure supported in one or
two dimensions but not the third. The gravitational time scale is generally
longer than the kinematic in this situation, 
so the large-scale collapse can occur later, in a chaotic
region that may have previously experienced several types of 
smaller scale instabilities. If the kinematic 
instabilities produce fine-scale 
structure, then the gravitational collapse pulls this structure together, 
forming large turbulent cores. 
A schematic diagram of this scenario is shown in Figure \ref{fig:teninter}.
Since the ridge is always unstable to collapse along its length, 
the condition for triggered cluster formation is the time when
the large-scale collapse becomes important, considering its growth and 
motion. This is approximately the time when the gravitational 
collapse rate equals
the inverse of the layer age. 

Doroshkevich (1980) and
Vishniac (1983) gave the instantaneous gravitational 
collapse time for a {\it constant-velocity} 
plane-parallel layer bounded by a shock on one side:
\begin{equation}t\sim\left(2\pi G\rho_{layer}\right)^{-1/2}
\sim\left(2\pi G\rho_0{\cal M}^2\right)^{-1/2}
\end{equation}
for preshock density $\rho_0$ and ${\cal M}=v_{shock}/c_{layer}$. 
This result came from the collapse rate $\omega$ where
\begin{equation}
\omega^2=2\pi G\sigma k-k^2c^2;    
\label{eq:vis}
\end{equation}
$\sigma$ is the mass/area in the layer
and $k=2\pi/\lambda$ is the wavenumber for wavelength $\lambda$.
For a 
steadily moving layer, we set the instantaneous collapse rate
equal to the inverse of the time,
and we set the net accumulated column density equal to $\sigma=
\rho_0vt$, to get the overall collapse time of the fastest-growing
wavelength:
\begin{equation}t\sim\left(\pi G\rho_0{\cal M}\right)^{-1/2}.
\end{equation}

\begin{table}
\caption{Triggered Clusters in Massive Cloud Cores}
\begin{center}\scriptsize
\begin{tabular}{ll}
Region&Reference\\
\tableline
M17&Thronson \& Lada 1983; Rainey et al. 1987;\\
&Greaves, White \& Williams 1992\\
Per OB2&Sargent 1979\\
Ceph OB3&Sargent 1979\\
Rosette&Phelps \& Lada 1997\\
S155/Ceph B cloud&Testi et al. 1995\\
G5.48-0.24&Koo et al. 1996\\
NGC 7538&Campbell \& Persson 1988\\
Sco-Cen&Blaauw 1991; Walter et al. 1994\\
Orion OB 1&Blaauw 1991\\
NGC 1962-65-66-70 in LMC&Will et al. 1995\\
NGC 346 in SMC&Massey, Parker \& Garmany 1989\\
\end{tabular}
\end{center}
\end{table}


For a {\it decelerating} layer, 
the growing dense core migrates to the shock front 
because it decelerates less rapidly than the lower density parts. 
At the shock front it can {\it erode} because of the transverse flow 
driven by the resulting shock curvature.
Erosion competes with growth by self-gravity so
the overall collapse can be {\it delayed} to the longer time
(Elmegreen 1989; Nishi 1992):
\begin{equation}t\sim0.5\left(G\rho_0\right)^{-1/2}.
\label{eq:e89a}
\end{equation}
This solution again 
came from a collapse rate $\omega\sim1/t$, but now		
\begin{equation}
\omega^2\sim2\pi G\sigma k-0.5 k^2v_{shock}^2.    
\label{eq:e89}
\end{equation}
Note the substitution here of $v_{shock}$ for $c$ in equation
(\ref{eq:vis}) (Elmegreen 1989). 
Other processes leading to internal shear may delay the collapse to this
timescale too (Doroshkevich 1980).
Equation (\ref{eq:e89}) 
for $\omega$ indicates that collapse occurs
only when the layer is so strongly self-gravitating that forced 
transverse
motions behind the shock front
cannot disrupt the density perturbations.

For a decellerating layer, collapse on the short timescale
$\left(\pi G\rho_0{\cal M}\right)^{-1/2}$ for ${\cal M}>>1$ leads 
to small clumps that may not be self-gravitating
(Lubow \& Pringle 1993), but if cooling is sufficient and
self-gravitating clumps do form, then any stars
they produce will move out of the front of the layer and leave.
This is because the deceleration of the layer, $\sim v_{shock}/t$,  
always
exceeds the self-gravitational
acceleration perpendicular to the layer, $G\sigma$, 
for timescales $t<<(G\rho_0)^{1/2}$
(Elmegreen 1989; Nishi 1992). 
To see this, we set $\sigma\sim v_{shock}\rho_0t$,
then $v_{shock}/t>>G\sigma$ whenever $t<<(G\rho_0)^{1/2}$. 
The observation of giant condensations {\it inside}
and gravitationally bound to 
swept-up layers (Table 2), and of the ages and dimensions of these 
layers for molecular cloud densities exceeding $10^3$ cm$^{-3}$, 
are consistent with the {\it long} time scale of
$\sim0.5\left(G\rho_0\right)^{-1/2}$ for triggered clusters in 
swept-up gas. 
This implies that gravitational rather than kinematic instabilities
govern the onset of cluster formation in a shocked layer. 
Any observation of moving young stars ahead of the swept-up layer, 
inside the unshocked molecular cloud, would be evidence for 
triggering on the short time scale.

The analyses of layer instabilities by Voit (1988) and
Whitworth et al. (1994a,b)
treated primarily the first significant collapse of the layer, on
the short timescale, $(G\rho_0{\cal M})^{-1/2}$, when the layer is
still primarily pressure-bound. They did not consider
the forward migration and erosion of condensations, shear 
behind the shock, and the
possibility that stars migrate individually out of the
front of the layer if the layer decelerates. Nevertheless, 
their assumptions are good 
for non-decelerating layers, as might occur between
two colliding clouds, and this was one of the applications 
considered by Whitworth et al..

Lubow \& Pringle (1993) also considered the collapse of a
highly compressed layer, and explained analytically 
the result found in 
Elmegreen \& Elmegreen (1978b), that the layer 
has an early gravitational instability with a wavelength
comparable to the layer thickness and a growth rate comparable
to the internal gravitational timescale, $(G\rho)^{-1/2}$ for 
internal layer density $\rho$. This is true 
even though the layer is confined more
by pressure than self-gravity in the perpendicular direction. 
The reason they gave is that the initial fast collapse is an incompressible
deformation of the surface under these high-pressure conditions,
and such deformations can happen quickly. The deformation is
curved symmetrically about the center without deceleration, 
and so the boundary pressure is directed inward, aiding 
the self-gravity (see also discussion in Elmegreen 1989).
Compressive gravitational instabilities on a larger scale, which may
give rise to cluster formation, grow later. 
Lubow \& Pringle (1993) considered applications
to cloud collisions and concluded that substantial cooling is required
to trigger the fast collapse into stars (as discussed also by 
Hunter et al. 1986).

Another discussion of the collapse of decelerating layers
was by Nishi (1992), who explained the linearized results obtained
numerically by Elmegreen (1989) in an analytical fashion, considering
appropriate simplifying assumptions. 
Nishi (1992) found the short oscillating pressure-driven 
solutions and the long,
monotonic, gravitationally-driven solutions, and pointed out the 
importance of a dimensionless parameter ($\alpha)$, 
which is essentially the
ratio of the Jeans length to the scale height
inside the layer, or, similarly, the ratio of the deceleration
of the layer to the self-gravitational acceleration perpendicular
to the layer. 
A layer begins its evolution with a large value
of $\alpha$ and ends with a small value as self-gravity becomes
important. Nishi (1992) found that the monotonic gravitational collapse
begins when $\alpha\sim1$, and this corresponds to the long
overall time scale, $\sim(G\rho_0)^{-1/2}$ for external (pre-shock)
density $\rho_0$, as given by equation (\ref{eq:e89a}) above. 

A 2D hydrodynamic simulation of the collapse of a decelerating isothermal
layer
was done by Yoshida \& Habe (1992). They confirmed the oscillatory behavior
at small wavelength and the monotonic collapse at large wavelength 
that was
found by others, they got the predicted migration of the dense condensation 
to the front of the layer, and they got the expected 
timescale for gravitational
collapse, i.e., 
comparable to about $0.5\times$ the free fall time in the unshocked
gas.  They found several new effects however.  
At moderately long wavelengths where gravity begins to be important, 
the tangential flow behind the shock front that was found by Vishniac (1983)
and others to cancel the pinch force from pressure there, was
compensated in the 2D calculation by an opposing tangential force
far behind the shock front in the dense gas, driven by this pinch
force and by self-gravity. Because of this, a dense condensation could
form at a leading perturbation and not be eroded away, as found
by Elmegreen (1989) in the linear analysis, and so the condensation
stayed at the leading perturbation and grew to significant densities, 
always with a slight protrusion out of the front. After a time equal to
half the external free fall time, the condensation had an escape
velocity larger than the transverse flow speed, so the erosion never
got to be effective.  At longer wavelengths, gravity is even more
important and the collapse, still monotonic and unperturbed by erosion, 
occurs without much migration to the front of the layer. This is
because the growth at longer wavelengths is slow enough that when it
becomes important, the layer has slowed down so much that its deceleration
is significantly less than the perpendicular 
gravitational acceleration toward the
midplane. They also obtained a minimum wavelength for gravitational
collapse equal to about the layer Jeans length, $2\pi G\sigma/c^2$,
for mass column density $\sigma$ and velocity dispersion $c$, 
evaluated at the time when the deceleration of the layer equals
the self-gravitational acceleration perpendicular to the layer
($\alpha=1$ in the notation of Nishi 1992). 
Under typical conditions, the mass contained in this minimum wavelength
is several hundred solar masses or more, suggesting again the
formation of embedded clusters rather than individual stars. 
These results were for an initial corrugation of fixed
size; the results for other perturbations and for colder postshock
gas are not known.  The results with magnetic fields are also not
known from any of these analyses. 

It is interesting that the short 
and long timescale solutions for layer collapse
were discussed in the Russian literature before these ideas surfaced
in the West.  
Doroshkevich (1980) used the same shock boundary condition 
as Vishniac (1983) and got oscillatory solutions for unstable growth on 
the short time scale, as Vishniac and others did  
(see equations 7a and 10 in Doroshkevich), and he 
included shear flow behind the shock and got the long time scale for
gravitational collapse as in Elmegreen (1989)
for the analogous symmetric mode (i.e. collapse along the layer; see equations
19 or 26 in Doroshkevich and note that his $\beta$ is our ${\cal M}$). 
Doroshkevich (1980) summarized his results with the prescient statement:
"Shear flows therefore sharply diminish the maximum
growth rate of the symmetric perturbation mode. The wavelength corresponding to
the maximum growth rate will increase very rapidly. At
the same time, however, a hydrodynamic instability will set in,
developing within a comparatively short time interval on small scales
and causing the layer to break up into small clouds. Later 
the clouds might clump together through the
action of gravitational instability that has
developed on scales of long wavelength. These
results depend only weakly on the choice of velocity profile (provided the
profile is reasonably smooth)." 
This is pretty close to what we think today. Doroshkevich (1980) applied
his results to cosmological pancakes; here they are relevant to the formation
of star clusters in compressed layers at 
the edges of HII regions and elsewhere. 

\begin{figure}
\vspace{2.1in}
\includegraphics{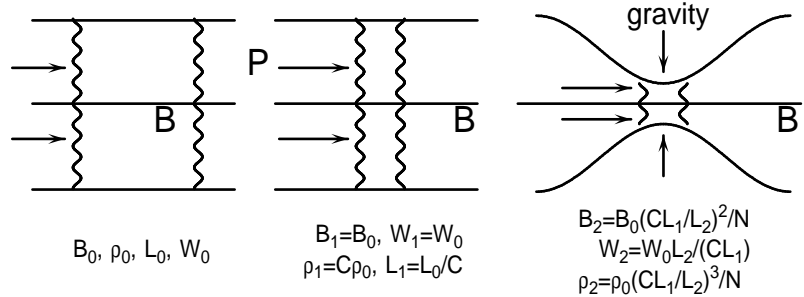}
\caption{\it Diagram of the collapse of a layer that moves parallel
to the mean magnetic field orientation. The initial values are
denoted by subscript 0; values after a purely lateral
compression are denoted by 1, and after the gravitational
collapse by 2.  The result is written in terms of
$L_1/L_2\approx 1$.  After the collapse,
the density in the cloud has increased by the
cube of the lateral compression factor $C$.
}
\label{fig:tenpar2}
\end{figure}

What is the actual mechanism of star formation in this scenario?
All that the instability analysis describes is the collapse of a layer into 
one or more dense cores. 
In fact, the collapse is probably first to filaments, and then
to cores inside the filaments (Miyama, Narita, \& Hayashi 1987).
Presumably the mechanism of star formation is the same as in any
other dense core once it forms, but faster in this case because of the
higher density following the compression.
This higher density can be quite large for initial compression 
parallel to the magnetic field, because it increases not only from
the direct one-dimensional compression caused by the HII region, but
also because of the subsequent collapse 
of the compressed gas perpendicular to the field.
A schematic diagram of this two-step process is shown in Figure
\ref{fig:tenpar2}.
The core density divided by the preshock density 
equals approximately the {\it cube} of the average
density compression factor in the layer alone (Elmegreen 1985, 1992). 
Most of this density increase
is from self-gravity;
only a small amount
is from the
external pressure.

In Figure \ref{fig:tenpar2}, the initial cloud
width is $W_0$ perpendicular to the field,
the length is $L_0 $ along the field, the density is $\rho_0$, and
the magnetic field strength $B_0 $.
After the compression (subscript 1),
the width and field strength are the same, $W_1 = W_0 $,
$B_1 = B_0 $, the density is higher by the
shock-compression factor $C$, $\rho_1 = C \rho_0$, and
the length is smaller by the inverse of this
factor, $L_1 = L_0 / C$.

The final configuration (subscript 2) considers that the both the initial
and final clouds are in equilibrium with the magnetic field 
resisting self-gravity. This requires magnetic field strengths
proportional to the transverse column densities:
\begin{equation}
B_0 = K \rho_0 W_0 \;\;\;\;\;;\;\;\;\;\;
B_2 = K \rho_2 W_2. 
\end{equation}
Then with mass and magnetic flux conservation, the final density
becomes
\begin{equation}
{{ \rho_2} \over { \rho_0}}= {1\over {N}}\left(
{{ CL_1} \over { L_2}} \right)^3 \end{equation}
for $N$ cores. 
We expect $L_2 < L_1$ because of the extra gravity
in the collapsed core,
and typically $N$ for the formation of OB subgroups 
is in the range 1-5; then the compression
factor after the collapse can be comparable to $C^3$, which may be a
factor of 100 to 1000. 

In the collect-and-collapse scenario, several  
dense fragments form in a ridge of swept-up gas
because of gravitational instabilities. 
Stars form in these fragments at
a much higher rate than they would have formed in the same gas
without the compression because
of the higher fragment density. 
The time scale for the whole triggering process is about the time scale
for the layer to collapse, as given above by equation \ref{eq:e89a}, 
because the star formation
time inside each fragment should be relatively short compared to this,
such as $(G\rho_2)^{-1/2}$. 

The important aspect of this scenario is that there is a {\it delay}
between the expansion of the HII region and the 
onset of
triggered star formation in the swept-up layer. 
There is also bulk {\it motion} of the compressed gas and
triggered clusters. 
The delay results from the
layer's initial stability {\it against} collapse that comes from the
shock-forced transverse motions, erosion, and turbulence.  This is
evident from the term with $v_{shock}$ in equation \ref{eq:e89}. 
A similar point was made by Vishniac (1983), 
but here we stress that
the gravitational instability should not be {\it prevented} by the internal
motions, but only {\it delayed}. As a result, all of the stars in a
triggered cluster at the edge of an OB association should be significantly
younger than the stars in the shock-driving cluster. That is, star
formation should not be continuous in the layer, beginning with the onset of the
expansion, but delayed. Then, when it begins, it happens rapidly, so 
all the triggered stars have about the same young age. 
Also because of this, there is
not a smooth distribution of intermediate age stars between the old
cluster and the young cluster, but essentially no intermediate age
stars between the two clusters. 
These two aspects of the collect-and-collapse scenario are
in agreement with observations of OB subgroups (Blaauw 1964, 1991).  

In contrast, triggering in 
the globule-squeezing scenario is
immediate because the pre-existing clumps are compressed quickly.
Such triggering also takes place throughout the HII region
because residual cloud pieces can be 
anywhere.
Globule compression differs also 
because it probably forms individual stars or
small stellar systems, while the 
collect-and-collapse scenario forms whole clusters 
if the compressed layer contains enough mass. 

Other processes that can 
structure dense molecular gas and trigger star formation are likely to 
occur as well. 
For example, the forced motion of the ridge can cause 
pre-existing clumps in the
moving part of the cloud to collide and coalesce with pre-existing
clumps in the unshocked part of the cloud.  
Such collisions can lead to star formation on a clump-by-clump basis
(Greaves \& White 1991).
In addition, the 
pre-existing clumps cause irregularities in the shock front,
and these lead to
transverse motions and the accumulation of large cloud
fragments in the ridge, even without self-gravity
(Kimura \& Tosa 1993).
The resulting fragments have the appearance of bright rims, and may
contain star formation at enhanced rates when gravity becomes important
(Elmegreen, Kimura, \& Tosa 1995).  

\subsection{Large Scale Triggering: Shells and Rings}

Expansion around a centralized pressure source leads to a shell or ring
that can become gravitationally unstable to form clouds and new stars
along the
periphery. The actual physical size of the shell can be large or small, 
depending on the pressure and external density.  
If the external density is high, then the shell and propagation distance 
will be small. For expansion into the ambient medium, which generally has
a low density, the shell can be several hundred parsecs in size. 
Examples of giant shells with young clusters along the periphery are
found in our Galaxy and neighboring galaxies, such as 
M31 and the LMC,
as summarized in Table 3.  
A review of giant shell formation and triggered star formation is in 
Tenorio-Tagle \& Bodenheimer (1988). 

After Heiles' (1979) discovery of giant HI shells in our Galaxy, the
first detailed model for their origin around high pressure
OB associations was 
by Bruhweiler et al. (1980).
Other models for shell formation 
soon followed, including one in which 
an extragalactic cloud collides with the galaxy disk 
and the resulting shell fragments into giant molecular
clouds (Tenorio-Tagle 1981).  
Propagation of star formation over 
such distances had already been assumed by Gerola \& Seiden (1978).
Applications of the shell-collapse scenario 
to the Orion, Perseus and Sco-Cen clouds
in the Lindblad ring
were then made independently by Olano (1982) and Elmegreen (1982),
with the first detailed summary of 
observed regions in Elmegreen (1985). 

Analytical work on gravitational instabilities in expanding
shells began with Ostriker \& Cowie (1981) and Vishniac (1983). An early
computer simulation for the collapse of the Lindblad ring 
was in Elmegreen (1983, 1985).
Consideration of both the theory for shell expansion and the
theory for collapse 
was in McCray \&
Kafatos (1987), who used a model of shell formation in which the
energy continuously increased, as if by a steady wind or continuous
supernovae.  Differential rotation 
was introduced by Tenorio-Tagle \& Palous (1987) 
and Palous, Franco, \& Tenorio-Tagle (1990).

\begin{table}
\caption{Triggered Star Formation in Swept-up Shells or Rings}
\begin{center}\scriptsize
\begin{tabular}{ll}
Region&Reference\\
\tableline

$\;\;\;\;\;\;\;\;${\it Galactic Clouds}&\\
Lindblad's Ring&Elmegreen 1982; Olano 1982; Taylor, Dickman,\\ 
&\& Scoville 1987; Franco et al. 1988; Comer\`on,\\ 
&\& Torra 1994b\\
W3/W4&Lada et al. 1978; Dickel et al. 1980; Thronson,\\ 
&Lada, \& Hewagama 1985; Routledge et al. 1991;\\ 
&Digel et al. 1996; Tieftrunk et al. 1997\\
W5&Sato 1990\\
IC 443, W28, W44, S147, HB21&Odenwald \& Shivanandan 1985\\
Ara OB1 field&Rizzo \& Bajaja 1994\\
W75&Ward-Thompson \& Robson 1991\\
Cygnus superbubble&Comer\`on \& Torra 1994a\\
MonR2&Hughes \& Baines 1985; Xie \& Goldsmith 1994\\
NGC 1333&Langer, Castets, \& Lefloch 1996; Warin et al. 1996\\
GS235-02&Jung, Koo, \& Kang 1996\\
G24.6+0.0&Handa et al. 1986\\
$\;\;\;\;\;\;\;\;${\it Magellanic Clouds}&\\
Constellation III&Dopita, Mathewson \& Ford 1985\\
LMC2&Wang \& Helfand 1991\\
LMC4&Domg\"orgen, Bomans \& De Boer 1995; Olsen et al.\\
&1997\\
N11&Walborn \& Parker 1992; Parker et al. 1992;\\
&Rosado et al. 1996\\
DEM 152 in N44&Oey \& Massey 1995\\
NGC 2214&Bhatt \& Sagar 1992\\
$\;\;\;\;\;\;\;\;${\it Galaxies}&\\
M33&Deul \& den Hartog 1990; Palous 1991\\
M31&Brinks \& Bajaja 1986; Brinks, Braun, \& Unger 1990\\
IC 2575&Martimbeau, Carignan, \& Roy 1994\\
NGC 1313&Ryder et al. 1995\\
Ho II& Puche et al. 1992\\
BCD SBS 0335-052&Thuan, Izotov \& Lipovetsky 1997\\
NGC 1620 &Vader, \& Chaboyer 1995\\
Giant Extragalactic HII regions&Mayya \& Prabhu 1996\\
\end{tabular}
\end{center}
\end{table}

Applications of long-range propagating star formation 
to galactic structure and
spiral arm formation were made by many others following
Mueller \& Arnett (1976) and
Gerola \& Seiden (1978).
A recent review of their work is in 
Seiden \& Schulman (1990), and other recent work is in 
Korchagin \& Riabtsev (1992),
Newkirch \& Hesse (1993),
Jungwiert \& Palous (1994),
and Palous, Tenorio-Tagle, \& Franco (1994).
These latter papers consider asymmetric propagation, as might
result in the presence of galactic shear. 

The criterion for star formation in an expanding shell or ring
is analogous to that in the
collect-and-collapse scenario discussed in the previous section. 
Shells and rings behind shock fronts contain a variety
of instabilities, driven by kinematic, magnetic, and
self-gravitational processes. Some of these stir up the gas and
create small scale turbulence and structure 
not related to star formation.
Gravitational instabilities lead to the formation of large condensations
inside the swept-up material, and some of these 
may produce embedded clusters. 
Thus the criterion for the onset of cluster formation in a shell or
ring is the time at
which gravitational instabilities become important.  

The divergence of expanding shells or rings gives them an initial
stability
against self-gravitational collapse. 
For continuous energy deposition into a shell, so that
$v_{shock}\propto t^{-0.4}$ (Castor et al. 1975), the shell is first
unstable when (Elmegreen 1994c)
\begin{equation}t\sim\left(G\rho_0{\cal M}\right)^{-1/2},
\end{equation}
and the growth rate equals $1/t$ when 
\begin{equation}t\sim1.25\left(G\rho_0{\cal M}\right)^{-1/2}.
\label{eq:t1}
\end{equation}
In these equations, the 
preshock density is $\rho_0$ and the ratio of shock speed to
velocity dispersion in the swept-up gas is ${\cal M}=v_{shock}/c$. 
Slightly different results were obtained by Whitworth et al. (1994a) 
without the effects of shell divergence.
Theis et al. (1997) showed how the ratio of the fragmentation time
to the time of first instability is constant, and they also
showed that any density gradient in the surrounding medium 
has to be shallower than isothermal for the self-gravity 
of the accumulated matter to ever overcome the stability 
given by the expansion. 

For a ring, the growth rate $\sim1/t$ when (Elmegreen 1994c)
\begin{equation}t\sim1.5\left(G\rho_0{\cal M}^2\right)^{-1/2}.
\label{eq:t2}
\end{equation}
The extra power of $M$ for the ring
compared to the shell
comes from the different geometry.

Equations (\ref{eq:t1}) and (\ref{eq:t2})
give the collapse conditions that should 
signal the onset
of star formation. 
They include explicitly the 
self-gravity, internal pressure, and expansion of the region, while
the kinematic instabilities that are also present are 
absorbed into the internal velocity dispersion through
the term ${\cal M}$. 
These small scale instabilities increase $c$ and decrease
${\cal M}$, which lengthens the timescale for
collapse, as discussed in the previous section. 
Rings are more important than shells
for large scale expansion in galaxies
because most of the accumulated material is in the midplane.  
Indeed, 
Ehlerov\'a et al. (1997) show with a numerical simulation how
only the equator of a shell expanding into a thin galactic
disk becomes unstable. 

Another condition for star formation is that the swept-up layer has to 
be optically thick to starlight so it can cool and become dense. 
This condition was written by 
Franco \& Cox (1986) as $N>10^{21}(Z_\odot/Z)$ cm$^{-3}$ 
for total shell column density $N$ and metallicity $Z$. 
This second condition is 
implicitly included in the first 
through the term $M$.  If the swept-up region cannot cool, then
$M$ is small and the collapse takes a long time. Once cooling begins, 
$M$ increases and the collapse time drops. Thus cooling can trigger the
instability, as well as the continued accumulation of mass. 

If $\omega\sim1/t$ at the onset of cloud and star formation,
as assumed above, and if we write the variables in physically
realistic units with $n_0$ the ambient density ahead of
the front and $c$ the velocity dispersion inside the
swept-up layer, then for a shell:
\begin{eqnarray}t_{SF}\sim103\left({{n_0{\cal M}}\over{{\rm cm}^{-3}}}
\right)^{-1/2} \;\;{\rm My}\hfill\hfill\\
R_{SF}\sim176M^{1/2}\left({c\over{{\rm km}\;{\rm s}^{-1}}}
\right)\left({{n_0}\over{{\rm cm}^{-3}}}\right)^{-1/2} \;\;{\rm pc}.
\end{eqnarray}
and for a ring:
\begin{eqnarray}t_{SF}\sim124\left({{n_0{\cal M}^2}\over{{\rm cm}^{-3}}}
\right)^{-1/2} \;\;{\rm My}\hfill\hfill \\
R_{SF}\sim211\left({c\over{{\rm km}\;{\rm s}^{-1}}}
\right)\left({{n_0}\over{{\rm cm}^{-3}}}\right)^{-1/2} \;\;{\rm pc}.
\end{eqnarray}
These results indicate that a shell or ring has to be relatively large before
star formation appears along the periphery if the ambient density is
about 1 cm$^{-3}$,
the average for a galaxy disk at the Solar radius.
Inside molecular
clouds or in starburst regions, 
the density can be much larger, perhaps $10^3$ 
cm$^{-3}$ or more. Then the triggering time and shell radius can be
small. 

Once a ring 
or shell stops expanding, the collapse proceeds quickly.
Recall that for a moving ring, from equation \ref{eq:t2},
$t_{SF}\sim{{1.5}/{\left(G\rho\right)^{1/2}}}$
with $\rho=\rho_0{\cal M}^2$,
but, using the same analysis for a static ring, 
\begin{equation}t_{SF}\sim{{e^{0.5\left(1+c^2/G\mu_0\right)}}\over
{\left(4\pi G\rho\right)^{1/2}}}
\sim{{0.8}\over {\left(G\rho\right)^{1/2}}};
\end{equation}
$\mu_0$ is the mass/length of the ring
and $\rho$ is the density inside.
If the expansion history is unknown, then the SF time can
be assumed to be between these two limits, which corresponds to
\begin{equation}
t_{SF}\sim 66  -  124 \;\; n_{inside}^{-1/2} \;\;{\rm My}.
\end{equation}

\subsection{Star Formation Triggered by Cloud Collisions} 

The dense shocked gas between 
two colliding clouds is another region where gravitational 
instabilities can lead to triggered star formation. 
The direct observation of this effect has been difficult, 
however. 

The problem is that very few clouds are likely to be undergoing
a collision at any one time. 
Considering a cloud cross section $\pi R^2$, speed $v$, and
density $n$, the collision time is $\sim 1/(\pi R^2 vn)$
and the duration of each collision is $R/v$. 
This implies that the fraction of the time spent in the collision, 
or the fraction of all clouds in an ensemble that are currently
colliding, is $\pi R^3n$, which is about the volume filling factor
of the clouds. This filling factor is usually very 
low for both clouds and their clumps, i.e., less than 10\% for
diffuse clouds and less than 1\% for molecular clouds, so very
few clouds are likely to be involved in a collision at any one
time in a Solar neighborhood 
environment. The fraction can be much higher
in spiral density wave shocks where the clouds converge
(Kwan \& Valdes 1983; Roberts \& Steward 1987; Kenney \& Lord 1991).
However, in other regions,
cloud collisions are so rare that very few examples
have been found.
Some of the proposed regions are in Table 4.

The collapse of shocked layers between converging flows has been
studied both analytically and numerically. 
Many of the results are the same as in the case of a shocked
layer or shell so they are not repeated here. 
This section summarizes the recent calculations that deal specifically 
with gravitational collapse during cloud collisions.

Non-magnetic clouds typically destroy
each other during an off-axis collision at relative velocities
exceeding several times the internal sound speed
(Kahn 1955; Chieze \& Lazareff 1980;
Hausman 1981; Gilden 1984) or several times the
escape speed (Vazquez \& Scalo 1989), whichever is greater. 
This destruction occurs for two reasons: (1) 
the shock between the clouds is confined only
in the direction perpendicular to the contact plane, so 
high-pressure 
shocked material can squirt out parallel to this
plane, and (2) the non-overlapping portions of the clouds
do not feel the pressure from the shock and do not slow down
to a common speed.  

\begin{table}
\caption{Triggered Star Formation in Cloud Collisions}
\begin{center}\scriptsize
\begin{tabular}{ll}
Region&Reference\\
\tableline

NGC 1333&Loren 1976\\
LkH$\alpha$&Loren 1977\\
MBM 55&Vallee \& Avery 1990\\
DR 15, DR 20&Odenwald et al. 1990\\
G110-13&Odenwald et al. 1992\\
HVC+disk collisions&Lepine \& Duvert 1994\\
Sgr B2&Hasegawa et al. 1994\\
W49N&Serabyn, G\"usten, \& Schulz 1993\\
W75/DR 21&Dickel et al. 1978\\
Orion&Greaves \& White 1991;\\
&Womack, Ziurys, \& Sage 1993\\
IRAS 19550+3248&Koo et al. 1994\\
IRAS 2306+1451&Vallee 1995\\
\end{tabular}
\end{center}
\end{table}

Early calculations suggested that because of this destruction,
star formation requires
head-on collisions (Stone 1970), or
collisions between identical clouds (Gilden 1984), or
collisions between clouds that are almost unstable initially
(Lattanzio et al. 1985; Nagasawa \& Miyama 1987).
In such head-on collisions, or in the shocked layer between 
two unconfined gas streams
(Hunter et al.  1986; 
Whitworth et al. 1994a), the compressed gas collapses on
the time scale $(G\rho)^{-1/2}$ for compressed density $\rho$.
This can be very short if there is cooling and the 
compressed density is large (Pongracic et al. 1992). 

Now it is believed that even oblique collisions can be effective in triggering
gravitational instabilities and star formation in the compressed
layer (Usami, Hanawa, \& Fujimoto 1995).  
Velocity shear at the interface reduces the growth rate
of the gravitational instability, as discussed above, 
but it may not eliminate the
collapse altogether. 

Recent models suggest that star formation can also be triggered
by more general collisions if there is small scale structure. 
Habe \& Ohta (1992) found that for collisions between 
two different clouds in hydrostatic equilibrium, the large cloud
is disrupted by the processes discussed above, but 
the small cloud is compressed to trigger star
formation. This is because the small cloud behaves like a
cometary cloud, completely engulfed and confined by a large-scale gas flow.
An even more realistic case was considered by 
Kimura \& Tosa (1996), who found that collisions
between clumpy clouds trigger collapse in
the colliding clumps. 

Stars may also form during the collisions between clumps in a turbulent cloud.
This process is spontaneous in terms of the cloud evolution, that is, 
it does not need an {\it external} trigger for star formation, but
it is triggered in terms of the stability of each clump, i.e., each
clump needs the collision to form a star. 
A recent model using SPH of the gravitational instability of a 
compressed interface between two colliding clumps is in 
Pongracic et al (1992). They find that the instability operates much
faster than the gravitational timescale in the uncompressed 
clouds. It is probably more like $(G\rho)^{1/2}$ for compressed 
density $\rho$ (Whitworth et al. 1994a).  
This is consistent with the discussion 
in section \ref{sec:intermed} of the collapse
time in a non-decellerating layer.

Even without pre-existing clumps, supersonic turbulence inside clouds will make
dense sheet-like structures in the converging flows, and these structures
can be unstable to form stars by gravitational collapse, as in the clump
collision scenario. 
Early versions of this model were in Sasao (1973), with more recent work
in 
Sabano, \& Tosa (1985),
Elmegreen (1993), Padoan (1995), Vazquez-Semadeni, Passot, \& Pouquet
(1996), and Padoan, Nordlund \& Jones (1997).
The detailed physical processes are similar to those discussed by
Hunter et al. (1986) and others.
A new instability for 
oblique and shearing shock fronts at the interface between colliding
streams combines self-gravity with the Kelvin-Helmholtz
instability; it may give small scale structure and even gravitational
collapse into stars (Hunter, Whitaker \& Lovelace 1997). 

Models of clump or cloud collisions with magnetic fields are in their
infancy. Some first studies are in 
Byleveld, Melrose, \& Pongracic (1994)
and Byleveld, \& Pongracic (1996).

All of these models, whether they involve external shocks, 
converging flows, or clump collisions, make flattened gas sheets at
some point during the star formation process. Larson (1985) 
has emphasized this point, and Hartmann et al. (1994, 1996)
have considered the collapse of such layers and the spectra
of the resulting stars. 

\section{Turbulence Effects}

Turbulence changes our ideas about star
formation and triggering in several ways.  
For example, it makes clouds clumpy even without self-gravity, 
presumably making fractal structure in the gas. This gives the
clouds a
hierarchical nature, and may ultimately cause 
stars to form in clusters, associations and complexes. 

Pre-compression 
density and kinematic structures resulting from turbulence 
can also distort a shock front, causing
bright rims to form and the 
acceleration to be intermittent.
Positive acceleration makes
comet shapes, while 
negative acceleration makes bullets.

In a turbulent medium,  
the ambient pressure affects the degree of gravitational self-binding 
because the velocity dispersion of a virialized cloud depends on the 
pressure:
\begin{equation}
\Delta v_{virialized\;\;cloud} \sim G^{3/8} P^{1/8}M^{1/4}.
\end{equation}
When the velocity dispersion is very high,  
significantly greater than $\sim10$ km s$^{-1}$, 
OB stars cannot easily destroy the clouds in which they form, and then 
massive clusters can form with high efficiency, allowing themselves
to remain bound when the little remaining gas eventually leaves
(Elmegreen \& Efremov 1997). 
Even for lower velocity dispersions, greater cloud binding at higher
pressure should lead to a greater probability of forming a bound cluster.
This implies that triggered star formation in high pressure regions
may preferentially form bound clusters. 
For example, cloud collisions have been proposed to 
provoke the formation of globular clusters in the Large
Magellanic Clouds (Fujimoto \& Kumai 1997).
Globular clusters do indeed form in high pressure environments, such as  
early galaxy halos,
interacting galaxies and starbursts,
and 
occasional dwarf galaxy GMC cores, particularly in or near
high pressure HII regions (Elmegreen \& Efremov 1997).

Another effect of turbulence is the 
time-size correlation, $t\propto R^{1/2}$, which implies that
larger regions form stars for a longer time (Elmegreen \& Efremov 1996).
This observation may be 
related to the observed expansion of OB associations 
in the sense that every large region of star formation, which in this
case may be a whole OB association, contains smaller regions as part of the
natural hierarchical structure, and the large regions have larger
velocity dispersions at birth than the small regions. Thus there will
be a systematic progression in size, age, and velocity dispersion 
for all regions of star formation, including the progression from 
OB subgroups to whole associations. 
This will make the large scale association appear to be an expanded
version of the smaller scale subgroup, when in fact all may be parts of
a continuous hierarchy of structure that extends all the way from
small, multiple stellar systems to giant star complexes. 

\begin{figure}
\vspace{3.in}
\includegraphics{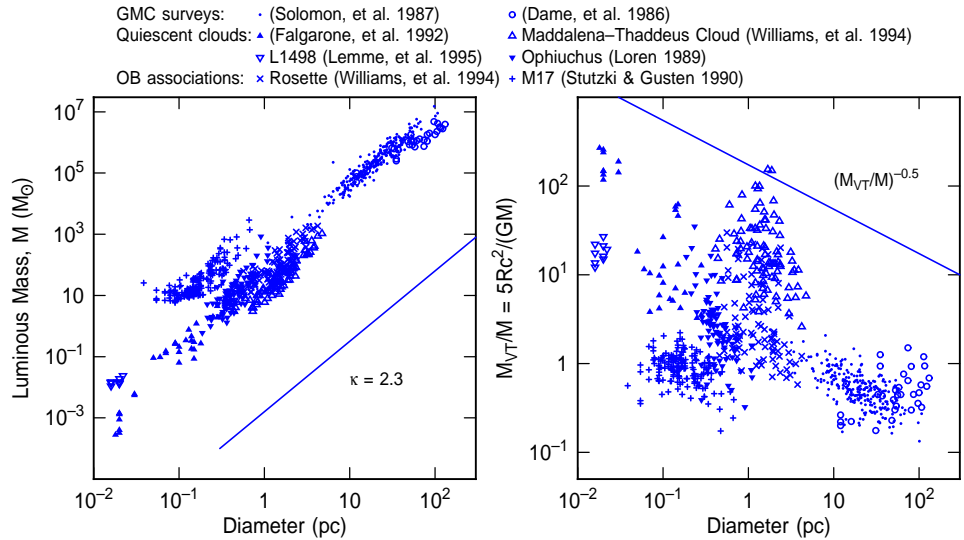}
\caption{\it The masses of molecular clouds and clumps are 
plotted versus their
FWHM diameters on the left, and the ratio of virial mass to luminous mass 
for the same clouds
is plotted versus diameter on the right. 
}
\label{fig:vt}
\end{figure}

The problem of star formation is considerably harder if much of the
small scale structure observed in clouds is from turbulence, as opposed to,
say, gravitational instabilities. 
Turbulent structures need not be self-gravitating,
and indeed many of the small clumps in giant molecular clouds appear to be
non self-gravitating. Then one wonders what transition takes place before they
form stars.  Perhaps collisions between turbulent clumps 
are {\it required} to trigger
star formation, as in the models discussed above. 
Vazquez-Semadeni, Passot, \& Pouquet (1996) discuss constraints on the
polytropic index for turbulence to induce the collapse of clumps.

Figure \ref{fig:vt} (right) shows the ratio of virial mass to luminous mass 
for clumps in several surveys, as indicated by the symbols. The virial
mass is taken to be $5Rc^2/G$ for FWHM radius $R$, Gaussian velocity
dispersion $c$, and gravitational constant $G$. Evidently, the smaller
clumps are less self-gravitating than the larger clumps, with a systematic
progression toward higher $5Rc^2/(GM)$ with lower $R$ (Bertoldi \&
McKee 1992; Falgarone, Puget, \& P\'erault 1992; 
Vazquez-Semadeni, Ballesteros-Paredes, \& Rodriguez 1997).
Considering that the
luminous mass is $M\propto cTR^2$ for temperature T, 
this ratio should be $\propto c/(RT)\propto 1/(R^{0.5}T)$ for the a
velocity-size trend
$v\propto R^{0.5}$. The solid line shows the predicted trend
for constant $T$. 

\section{Conclusions}

Star formation can be triggered by dynamical processes when high
pressures surround pre-existing clouds, when
high pressures accumulate dense cloudy material into layers that
collapse gravitationally into dense cores, when high pressures 
accumulate the ambient material into shells or rings, which then
collapse into cores, and when 
clouds, clumps, or turbulent streams collide. Examples of triggered
star formation in these
cases were given, as were the thresholds for the onset of star
formation.  The evidence suggests that 
dynamical triggering is widespread and operates over 
a large range of scales. 

The primary reason for the ubiquity of triggered star formation 
is that the gravitational timescale
in a molecular cloud, $(G\rho)^{-1/2}$, is shorter than the lifetime
of an O-type star, so dense molecular gas is commonly 
induced to collapse in the midst of 
high pressures from HII regions, stellar winds and supernova explosions.
Also, the gravitational timescale in the ambient disks of galaxies
is shorter than or comparable to the lifetime of an OB association, 
so whole clusters can make giant shells and new molecular clouds. 
In different regions, the balance between these basic time scales
may change, and then triggered star formation can have more or
less importance compared to spontaneous processes. 

\begin{quote}
\verb"\acknowledgements"
\end{quote}
Comments on the manuscript by F. Comer\`on 
and E. Vazquez-Semadeni are gratefully
acknowledged.

\end{document}